\documentstyle[prd,aps,preprint]{revtex}
\begin{document}
\draft
\title{Analytical results for the confinement mechanism in QCD$_3$}
\author{Walter Dittrich and Holger Gies\thanks{E-mail address: 
holger.gies@uni-tuebingen.de}}

\address{Institut f\"ur theoretische Physik\\
          Universit\"at T\"ubingen\\
   Auf der Morgenstelle 14, 72076 T\"ubingen, Germany}
\date{}
\maketitle
\begin{abstract}
We present analytical methods for investigating the interaction of two
heavy quarks in QCD$_{3}$ using the effective action approach. Our
findings result in explicit expressions for the static potentials in
QCD$_3$ for long and short distances. With regard to confinement, our
conclusion reflects many features found in the more realistic world of
QCD$_4$.
\end{abstract}
\narrowtext
\section{Introduction}
Gauge theory models in other than $3+1$ space--time dimensions have
been a center of interest for many years. In the present article we
want to explore QCD$_{3}$ and work out its similarities and
differences in comparison to QCD$_{4}$, the presumably correct
theory of strong interaction. Our investigation follows in many
respects the work of Adler and Piran \cite{1} in QCD$_{4}$. Their
starting point, the renormalization group approach \cite{2}, is,
however, bound to fail; there is simply no renormalization group in
QCD$_{3}$ nor is there a Callan--Symanzik $\beta$--function, etc.,
well--known attributes of QCD$_{4}$; our theory, QCD$_{3}$, is
super--renormalizable. However, the infrared problem, which we are
about to analyze, is shared by both theories. Now, since the
incompletely understood gluonic vacuum structure is at the heart of
both QCD$_4$ and QCD$_3$, we must rely on some more or less reasonable
effective action models. In Adler's case it is the {\em leading--log}
model while ours might be termed the {\em leading--root} model. The
latter will be represented by the one--loop effective action with
constant color magnetic background field. Among many interesting
features revealed in our work is the QCD$_3$ vacuum acting like a
dielectric medium, the elliptic shape of the confinement region and
its scaling properties and, finally, the behavior of the static
potential between two massive color test charges for large and small
distances. In Section II we present the essentials for the calculation
of the one--loop effective action. In Section III we focus on the
large distance (confinement) problem and give an expression for the
linearly rising potential plus correction terms. In Section IV we
treat the short distance domain and derive the classical formula for
the interaction of two static charges augmented, again, by correction
terms. Section V summarizes our findings.

\section{Computation of the one--loop effective Lagrangian}
Since much work has been invested in the calculation of effective
Lagrangians, we will use some short cuts to quickly reach our present
goal. The Lagrangian for a pure SU(N) gauge field theory is

\begin{equation}
{\cal L}=-\case{1}{4} F^a_{\mu \nu} F^{a\mu \nu} \label{1}
\end{equation}
\begin{equation}
\text{where }\quad F^{a\mu \nu}=\partial^\mu A^{a\nu}-\partial^{\nu}
A^{a\mu} +g\, f^{abc}A^{b\mu}A^{c\nu}\, . \label{2}
\end{equation}
Working in $d=3$ dimensions we list the following dimensions of
various quantities, some of them to be introduced at a later stage:
$[g]=[Q]=$$ [A^0]=m^{1/2}$, $[J^0]=m^{5/2}$.

Next, the gauge field is decomposed into

\begin{equation}
A^{a\mu}=A^{a\mu}_{\text{B}}+b^{a\mu}\, , \label{3}
\end{equation}
where $b^{a\mu}$ represents the fluctuating Yang--Mills field. The
external field configuration $A^{a\mu}_{\text{B}}$ that probes our QCD
vacuum is the commonly chosen color magnetic background field

\begin{equation}
A^\mu_{\text{B}}=-\case{1}{2} F^{\mu \nu}_{\text{B}} x_{\nu}\, , \quad
F^{\mu \nu}_{\text{B}}=\left( 
                         \begin{array}{ccc}
                            0 & 0 & 0 \\
                            0 & 0 & B \\
                            0 &-B & 0
                         \end{array} \right)\, , \label{4}
\end{equation}
\begin{equation}
\text{where } F^{a\mu \nu}_{\text{B}}=F^{\mu \nu}_{\text{B}}V^a ,\,
A^{a\mu}_{\text{B}} =A^{\mu}_{\text{B}}V^a ,\, (V^a)^2=1. \label{5}
\end{equation}
The magnetic field points along a fixed unit direction $V^a$ in color
space. Inserting the parametrization (\ref{3}) into the Lagrangian
(\ref{1}) we obtain

\begin{eqnarray}
{\cal L}=&-&\case{1}{4} F^{a\mu \nu}_{\text{B}}F^a_{\text{B} \mu
\nu}+\case{1}{2} b^{a\mu} \left[ g_{\mu \nu} (D^{ab\sigma}_{\text{B}}
D^{bc}_{\text{B} \sigma}) \right. \label{6}\\
 & -&\left. (D^{ab}_{\text{B} \nu} D^{bc}_{\text{B} \mu}) + g\,
f^{abc}F^b_{\text{B}\mu \nu}\right] b^{c\nu}+{\cal O}(b^3), \nonumber
\end{eqnarray}
\begin{equation}
\text{where} \quad D^{ab}_{\text{B} \mu}=\bigl( \partial_\mu
\delta^{ab} +g\, f^{acb}A^c_{\text{B} \mu} \bigr). \label{7}
\end{equation}
The Lagrangian (\ref{6}) together with the background gauge fixing
term,

\begin{equation}
{\cal L}_{\text{GF}}=\case{1}{\alpha} b^{a\mu} (D^{ab}_{\text{B} \mu}
D^{bc}_{\text{B} \nu}) b^{c\nu}\, , \label{8}
\end{equation}
connects us with the effective Lagrangian via the functional integral

\begin{eqnarray}
N' \exp \left[ \text{i} \int \text{d}^3x\, {\cal L}^{\text{eff}}
\right]&&\nonumber\\ 
=N \int {\cal D}b\, &\triangle_{\text{FP}}&\, \exp \left[ \text{i}\int 
\text{d}^3x\,
({\cal L} +{\cal L}_{\text{GF}}) \right] . \label{9}
\end{eqnarray}
Using the Feynman gauge, $\alpha=1$, we obtain

\begin{eqnarray}
{\cal L}+{\cal L}_{\text{GF}}=-\case{1}{2} B^2+\case{1}{2}b^{a\mu}
[&&g_{\mu \nu} (D^{ab\sigma}_{\text{B}} D^{bc}_{\text{B} \sigma})
\nonumber\\ 
 &&+2g\, f^{abc}F^b_{\text{B}\mu \nu}]b^{c\nu} . \label{10}
\end{eqnarray}
To compute the two terms in the square brackets of (\ref{10}) we
follow the author of ref.\cite{3} and so reproduce the result (pay
attention to signs!)

\begin{eqnarray}
{\cal L}+{\cal L}_{\text{GF}}=&-&\case{1}{2}B^2 + \sum_{T^b\in
C_{\text{V}}} \case{1}{2} b^{b\mu} (\partial^2 g_{\mu \nu}
-\partial_\mu \partial_\nu)b^{b\nu} \nonumber\\
&+&\sum_{\alpha} W^{\mu \ast}_\alpha (D^2g_{\mu
\nu}+2\text{i}gQ_{\alpha}F_{\mu \nu})W^\nu_\alpha, \label{11}
\end{eqnarray}
\begin{eqnarray}
\text{where}\, D^\mu=\partial^\mu
+\text{i}gQ_\alpha A^\mu_{\text{B}}&=&\partial^\mu-\case{\text{i}}{2}
gQ_\alpha F^{\mu \nu}x_\nu,\nonumber\\
 F_{\mu \nu}&\equiv& F_{\text{B}\mu \nu}. \label{12}\\
\text{and}\, C_{\text{V}}=\{ T^b |\, [V,T^b]=0 \}&,& V=V^aT^a, 
\nonumber
\end{eqnarray}
where the $T^a$'s denote the standart SU(N)--generators:
$[T^a,T^b]=\text{i}f^{abc}T^c$ , Tr$(T^aT^b)=\case{1}{2}\delta^{ab}$.
The remaining generators $\not\in C_{\text{V}}$ can be expressed in
terms of eigenvectors of $V$ with eigenvalues $Q_\alpha$. Their
fluctuating Yang--Mills field components $b^{a\mu}$ form certain
complex linear combinations $W^\mu_\alpha$ depending on the choice of
$V$ in color space.

At this point we perform a Wick--rotation so that all subsequent
calculations are done in Euclidean space:

\begin{eqnarray}
N'&\exp&\left[ \int \text{d}^3x\, {\cal L}^{\text{eff}}_{\text{E}}
\right] \nonumber\\ 
&=&N'' \int {\cal D}b\, \triangle_{\text{FP}} \exp \left[ \int
\text{d}^3x\, ({\cal L}_{\text{E}}+{\cal L}_{\text{EGF}}) \right]
\nonumber\\ 
&=&N''' \int {\cal D}W^\ast{\cal D}W\, \triangle_{\text{FP}} \exp
\left[ \int \text{d}^3x (-\case{1}{2}B^2 \right. \label{13}\\
&&\qquad \left. +\sum_{\alpha}W_{\alpha \mu}^\ast (D^2\delta_{\mu
\nu}+2\text{i} gQ_\alpha F_{\mu \nu})W_{\alpha \nu}) \right].
\nonumber 
\end{eqnarray}
Due to the fact that the fields $W^\ast$ and $W$ have to be treated
independently the Faddeev--Popov Determinant appears in the form

\begin{eqnarray}
\triangle_{\text{FP}}&=&\triangle_{\text{FP}}^{W^\ast}\triangle_{
\text{FP}}^W=\text{Det}(-D^2)\text{Det}(-D^2) \nonumber\\
&=&\exp \bigl[ 2\ln \text{Det}(-D^2) \bigr] =\exp \bigl[ 2\text{Tr} \ln
(-D^2) \bigr] \nonumber\\
&=& \exp \left[ \int \text{d}^3x\, 2\text{tr} \ln (-D^2) \right], \nonumber
\end{eqnarray}
Now we can write (\ref{13}) as

\begin{eqnarray}
N'&\exp&\left[ \int \text{d}^3x\, {\cal L}_{\text{E}}^{\text{eff}}
\right] \nonumber\\
 &=& N''' \exp \left[ \int \text{d}^3x\,
\bigl(-\case{1}{2}B^2+2\text{tr} \ln (-D^2) \bigr) \right]
\label{14}\\ 
 &&\times \int {\cal D}W^\ast {\cal D}W\, \exp \left[ \int
\text{d}^3x\, \sum_{\alpha} W^\ast_{\alpha \mu}(D^2\delta_{\mu \nu}
\right. \nonumber\\ 
&&\qquad \qquad \quad \qquad \qquad\left.+2\text{i}gQ_\alpha F_{\mu 
\nu})W_{\alpha \nu} \right]. \nonumber
\end{eqnarray}
Here we meet the standard functional integral

\begin{eqnarray}
\int {\cal D}W^\ast{\cal D}W\, &\exp& \left[ -\int \text{d}^3x\,
W_\mu^\ast M_{\mu \nu}W_\nu \right] \nonumber\\
&=&\exp \bigl[ -\text{Tr} \ln M_{\mu\nu} \bigr].\nonumber
\end{eqnarray}
Hence eq. (\ref{14}) takes the form

\begin{eqnarray}
N'&\exp& \left[ \int \text{d}^3x\, {\cal L}_{\text{E}}^{\text{eff}}
\right] \nonumber\\
&=&N''' \exp \left[ \int \text{d}^3x\, \bigl(-\case{1}{2}B^2+
2\text{tr}  \ln(-D^2) \right. \nonumber\\
&&\qquad \quad-\left. \text{tr} \ln(-D^2\delta_{\mu
\nu}-2\text{i}gQ_\alpha F_{\mu \nu}) \bigr) \right], \label{15}
\end{eqnarray}
from which we read off the expression for ${\cal
L}_{\text{E}}^{\text{eff}}$: 

\begin{equation}
{\cal L}_{\text{E}}^{\text{eff}}=-\case{1}{2}B^2+2\text{tr}
\ln(-D^2)-\text{tr} \ln (-D^2\delta_{\mu \nu}-2\text{i}gQ_\alpha
F_{\mu \nu}). \label{16}
\end{equation}
There are several ways to compute the various traces in this
expression. We prefer the $\zeta$--function regularization method
\cite{4}. The result is

\begin{eqnarray}
{\cal L}_{\text{E}}^{\text{eff}}=&-&\case{1}{2}B^2
+|gB|^{3/2}\nonumber\\ 
&&\times\case{1}{2\pi} \left[ 1-\frac{(\sqrt{2}-1)}{4\pi}
\zeta(\case{3}{2}) \right] \sum_{\alpha}|Q_\alpha |^{3/2} .
\label{17} 
\end{eqnarray}
In ref. \cite{3} and \cite{5} we found a proper--time calculation of
expression (\ref{16}). We agree with the result contained in ref.
\cite{3}. The numerical value of $\zeta(\case{3}{2})$ is
$\simeq$2.61238 so that the sign in front of $|gB|^{3/2}$ is indeed
positive. Hence ${\cal L}^{\text{eff}}$ takes a maximum
($V^{\text{eff}}$ a minimum) at a nonzero value of the background
field: 

\begin{equation}
|B_{\text{ex}}|^{1/2}=g^{3/2} \case{3}{4\pi} \left[ 1-
\frac{(\sqrt{2}-1)}{4\pi} \zeta(\case{3}{2}) \right] \sum_\alpha
|Q_\alpha |^2 . \label{18}
\end{equation}
The gauge invariant generalization of (\ref{17}) can be obtained by
the replacement $B^2\rightarrow$ $-\case{1}{2}F^{a\mu \nu}F^a_{\mu
\nu}$ $=(\bbox{E}^a)^2-(B^a)^2$ $={\cal F}$:

\begin{equation}
{\cal L}^{\text{eff}}=\case{1}{2}{\cal F}\left(1-\case{4}{3} \left|
\frac{{\cal F}}{\kappa^2} \right| ^{-\case{1}{4}} \right), \label{19}
\end{equation}
\begin{equation}
\text{where } \kappa^{1/2}=|B_{\text{ex}}|^{1/2}=\text{eq.}(\ref{18}).
\label{20} 
\end{equation}
So we obtain a gauge field condensate ${\cal F}=\kappa^2$ due to
radiative corrections, just like in four dimensions, which determines
the interesting features of the model. If we choose
SU(N=3) as our gauge group and the unit color vector pointing along
the three-direction we find $\kappa^{1/2}=0.37245\dots g^{3/2}$. 

At this stage it is important to point out that we are only dealing
with the lowest (first) order loop approximation.  Higher order loop
calculations will certainly modify the position and the shape of the
extremum, i.e., the value of $\kappa$. However, concordant with our
own approach there are other strong indications for gauge field
condensation as implied, e.g., by the so called average action method
advocated by the authors of ref. \cite{3a}.  Thus, we expect our {\em
leading--root} model to describe the exact QCD$_{3}$--vacuum
structure at least qualitatively accurate.

Furthermore, we assume that the form of (\ref{19}) also holds for
static fields which are slowly varying in space.

\section{Flux confinement and the heavy quark static potential}
In this section we study the statics of two massive test charges at
large distances. The approximation in which leading QCD$_3$ radiative
corrections are retained is given by the effective Lagrangian
(\ref{19}). Following Adler \cite{1} we write for the potential of 
static, infinitely massive test charges [$J^a_\mu =(J^a_0,0)$]

\begin{equation}
V_{\text{static}}=-\text{extr}_{A^a_\mu} \left\{ \int \text{d}^2x\,
({\cal L}_{\text{eff}}(A^a_\mu)-A^a_0J_0^a) \right\}. \label{21}
\end{equation}
Limiting ourselves to the case of quark--antiquark source charges we
set ($R=2a$)

\begin{eqnarray}
J^a_0&=&Q\hat{q}^a
[\delta^2(\bbox{r}-a\bbox{\hat{x}})-\delta^2(\bbox{r}+
a\bbox{\hat{x}})] \nonumber\\
&=& \hat{q}^aJ_0 , \label{22}
\end{eqnarray}
where $\hat{q}^a$ is the internal unit vector in color space.
Similarly we introduce scalar and vector potentials $\varphi$
and $\bbox{A}$ by writing

\begin{equation}
A^a_0=\hat{q}^a\varphi\quad ,\quad A^a_j=\hat{q}^aA_j . \label{23}
\end{equation}
Our variational problem can now be restated in the form

\begin{equation}
V_{\text{static}}=-\text{extr}_{\bbox{A},\varphi} \left\{ \int
\text{d}^2x\,  ({\cal L}_{\text{eff}}({\cal F})-\varphi J_0)
\right\}, \label{24} 
\end{equation}
where
\begin{equation}
{\cal F}=\bbox{E}^2-B^2\, ,\quad \bbox{E}=-\bbox{\nabla}\varphi\,
,\quad B=\epsilon_{ij} \partial_i A_j. \label{25}
\end{equation}
With ${\cal L}_{\text{eff}}({\cal F})$ given by (\ref{19}) we write
more explicitly

\begin{eqnarray}
V_{\text{static}}&&=-\text{extr}_{\bbox{A},\varphi} \nonumber\\
&& \left\{ \int \text{d}^2x\,
\left( \case{1}{2} {\cal F} \left( 1-\case{4}{3} \left| \frac{{\cal
F}}{\kappa^2} \right|^{-\case{1}{4}} \right) -\varphi J_0 \right)
\right\} \label{26}\\
&&=:\text{extr}_{\bbox{A},\varphi} \left\{ \int \text{d}^2x\, {\cal
L}_{\text{stat}} \right\}. \label{27}
\end{eqnarray}
Given ${\cal L}_{\text{stat}}$ we now can apply the Euler--Lagrange
equations which imply the field equations

\begin{equation}
\bbox{\nabla \cdot D}=J_0\quad ,\quad  \epsilon_{ij} \partial_i
E_j=0=\epsilon_{ij} \partial_i H\, , \label{28}
\end{equation}
\begin{equation}
\text{where } \bbox{D}=\epsilon\bbox{E}\, ,\quad H=\epsilon B\, ,\quad
\epsilon=1 -\left| \frac{{\cal F}}{\kappa^2} \right|^{-\case{1}{4}}.
\label{29} 
\end{equation}
The source--free equation (\ref{28}) for the magnetic field can be
satisfied by
\begin{equation}
\epsilon B^2=0. \label{30}
\end{equation}
Here we have to distinguish three cases:

\begin{eqnarray}
\text{(Ia)}&& B=0\quad ,\quad \bbox{E}^2>\kappa^2\nonumber\\
\text{(Ib)}&& B=0\quad ,\quad \bbox{E}^2<\kappa^2\label{31}\\
\text{(II)}&&\, \epsilon\, =0 \quad ,\quad B^2=\bbox{E}^2-\kappa^2
.\nonumber 
\end{eqnarray}
For short distances we expect Coulomb--like field configurations with
$\bbox{E}$ large and $B$ vanishing. Hence there should exist a finite
region containing the source charges for which (Ia) is satisfied. In
this domain we have reduced our original variational problem to a
problem in nonlinear electrostatics with a field strength dependent
dielectric constant:

\begin{eqnarray}
\bbox{\nabla \cdot D}=J_0& ,& \epsilon_{ij} \partial_i E_j=0 , 
\label{32}\\
\bbox{D}=\epsilon (E)\, \bbox{E}\, ,\quad \epsilon
(E)&=&1-\sqrt{\frac{\kappa}{E}}\, ,\quad E=|\bbox{E}| \in {\rm I\!R}.
\label{33}
\end{eqnarray}
Now, in Adler's {\em leading--log} model of QCD$_4$, it proved very
effective to work with a manifestly flux conserving quantity. So let
us likewise parametrize $\bbox{D}$ by introducing a scalar flux
function $\bbox{D}=\bbox{f} (\Phi)$. Without going into all the
details (and subtleties) of how to arrive at the explicit relation
between $\bbox{D}$ and $\Phi$, we just state the result:

\begin{equation}
\bbox{D}=\left( \frac{1}{2} \frac{\partial \Phi}{\partial y},
-\frac{1}{2} \frac{\partial \Phi}{\partial x} \right). \label{34}
\end{equation}
(The authors of ref. \cite{6} missed the important factor
$\case{1}{2}$.) The boundary conditions imposed on the flux function
$\Phi (x,y)$ are

\begin{eqnarray}
&&\Phi (|x|<a, y\rightarrow 0)=Q,\nonumber\\
&&\Phi (x>a, y\rightarrow 0)=0,\label{35}\\
&&\Phi \rightarrow 0 \quad \text{for} \quad x^2+y^2 \rightarrow
 \infty. \nonumber
\end{eqnarray}
For future calculations it is useful to derive some relations between
the fields $E$ and $D$. From eqs. (\ref{33}) and still considering the
branch (Ia), where $E>\kappa$, we obtain

\begin{equation}
D=|\bbox{D}|=|\epsilon ||\bbox{E}|=\epsilon E\, ,\quad \epsilon
=\sqrt{1-\frac{\kappa}{E}} >0. \label{36}
\end{equation}
This equation implies $D=-\sqrt{\kappa}E^{1/2}+E$ or, setting
$e:=\sqrt{E}$, 

\begin{equation}
e_{1,2}=\frac{\sqrt{\kappa}}{2} \pm \sqrt{\frac{\kappa}{4} +D}\, .
\label{37}
\end{equation}
\begin{displaymath}
\text{We need } \quad e_{1,2}^2=\frac{\kappa}{2} \pm
\sqrt{\frac{\kappa^2}{4}+ \kappa D}
\end{displaymath}
and select the positive sign to guarantee a single--valued potential.
So we have the relation

\begin{equation}
E=E(D)=\frac{\kappa}{2}+\sqrt{\frac{\kappa^2}{4}+ \kappa D} + D.
\label{38}
\end{equation}
At last we turn to the solution of eq. (\ref{32}). In order to obtain
more insight into the behavior of the solution, we begin by rewriting
the field equation for $\Phi (x,y)$ in its characteristic form. So let
us start with

\begin{displaymath}
\epsilon_{ij} \partial_i E_j=0
\end{displaymath}
\begin{eqnarray}
\text{or }\quad 0&=&\partial_x E_y-\partial_y E_x= \partial_x \left(
\frac{D_y}{\epsilon} \right) -\partial_y \left( \frac{D_x}{\epsilon}
\right) \nonumber\\
&=&\partial_x \left( \frac{D_y \sqrt{E}}{\sqrt{E}
-\sqrt{\kappa}} \right)-\partial_y  \left( \frac{D_x \sqrt{E}}{\sqrt{E}
-\sqrt{\kappa}} \right) . \label{39}
\end{eqnarray}
Here we employ eq. (\ref{37}) and introduce the flux function via
relation (\ref{34}). The result of a rather lengthy chain of partial
derivatives is given by the exact field equation

\begin{equation}
\left( \partial^2_x +\partial^2_y +(\alpha -1)\partial^2_n \right)
\Phi=0, \label{40}
\end{equation}
\begin{displaymath}
\text{where }\quad \partial_n =\bbox{\hat{n} \cdot \nabla}\, ,\quad
\bbox{\hat{n}} =\frac{\bbox{\nabla} \Phi}{|\bbox{\nabla} \Phi |}
\end{displaymath}
\begin{equation}
\text{and } \alpha=- \frac{1+\frac{\kappa}{2}
\frac{1}{\sqrt{\frac{\kappa^2}{4} +\kappa D}}}{-\frac{\kappa}{2}
\frac{1}{D} -1 -\frac{\sqrt{\frac{\kappa^2}{4}+ \kappa D}}{D}} .
\label{41}
\end{equation}
Letting $\partial_l$ be the tangential derivative, we can replace the
coordinate derivatives ($\partial_x, \partial_y$) in (\ref{40}) by

\begin{displaymath}
\partial^2_x+\partial^2_y =\partial^2_l +\partial^2_n +\text{first
derivative terms} 
\end{displaymath}
and so the field equation (\ref{40}) takes the form

\begin{equation}
(\partial^2_l +\alpha\, \partial^2_n) \Phi + {\cal
O}(\partial_l,\partial_n)=0. \label{42}
\end{equation}
Now, for weak fields, $D \rightarrow 0$, i.e., away from the charges,
we have

\begin{equation}
\lim_{D  \to 0} \alpha = \lim_{D \to 0} \frac{\frac{\kappa}{2}
\frac{D}{\sqrt{\frac{\kappa^2}{4}+\kappa D}} +D}{\frac{\kappa}{2} +D+ 
\sqrt{\frac{\kappa^2}{4} +\kappa D}}=0 , \label{43}
\end{equation}
and for strong fields, $D \rightarrow \infty$, i.e., close to the
charges, we obtain

\begin{equation}
\lim_{D\to \infty} \alpha=1 . \label{44}
\end{equation}
The flux equation (\ref{40}) is quite similar to the one found in
QCD$_4$; it is of degenerating elliptic type and has a real
characteristic at a surface of constant $\Phi$, where $\bbox{\nabla}
\Phi =0$. Using the same arguments as in Adler's {\em leading--log}
model, we have here the first indication of confinement in QCD$_3$.
Next we want to show quantitatively that, in fact, the total flux
between two massive color charges is confined to a domain with a
characteristic as boundary on which $\Phi$ vanishes. To do so it is
useful to reformulate our problem in still another form for the
equation for $\Phi$. For this reason let us go back to eq. (\ref{32}).
We also know that $E$ depends on $D$ in a way stated in (\ref{38}):

\begin{displaymath}
f(D) := E(D)=\frac{\kappa}{2} + \sqrt{\frac{\kappa^2}{4} +\kappa D}
+D .
\end{displaymath}
Then we obtain

\begin{eqnarray}
0&=&\partial_x E_y-\partial_y E_x= \partial_x\left( \frac{E_y}{E}
f(D) \right) -\partial_y\left( \frac{E_x}{E} f(D) \right) \nonumber\\ 
&=&\partial_x\left( \frac{D_y}{D} f(D) \right)-\partial_y\left(
\frac{D_x}{D} f(D) \right) . \nonumber
\end{eqnarray}
Recalling relation (\ref{34}) we get

\begin{eqnarray}
\partial_x \left[ \frac{\partial_x \Phi\, f(D)}{((\partial_x \Phi)^2+ 
(\partial_y \Phi)^2)^{1/2}} \right]&& \nonumber\\
\mbox{} + \partial_y \left[
\frac{\partial_y \Phi\, f(D)}{((\partial_x \Phi)^2+ (\partial_y
\Phi)^2)^{1/2}} \right]& =&0, \label{45}
\end{eqnarray}
\begin{displaymath}
\text{where } \left( (\partial_x \Phi)^2 + (\partial_y \Phi)^2
\right)^{1/2} =2D.
\end{displaymath}
After performing the various partial derivatives we end up with the
following, still exact, differential equation for $\Phi$:

\begin{eqnarray}
0=&&\left[ (\partial_x \Phi)^2+ \left( \frac{g(D)}{D}+1
\right)(\partial_y \Phi)^2 \right] \partial_x^2 \Phi \nonumber\\
&+&\left[ \left( \frac{g(D)}{D}+1 \right)(\partial_x \Phi)^2+
(\partial_y \Phi)^2 \right] \partial_y^2 \Phi \nonumber\\ 
&-&2\frac{g(D)}{D} \partial_x \Phi\, \partial_y \Phi\, \partial_{xy}
\Phi , \label{46}
\end{eqnarray}
\begin{equation}
\text{where } g(D):= \frac{f(D)}{f'(D)} -D . \label{47}
\end{equation}
At this stage we make contact with calculations contained in ref.
\cite{7}. Needless to say, a solution of (\ref{46}) is not easily
available. However, being interested in the far--field approximation,
we now study the limiting case of weak fields:

\begin{eqnarray}
f(D)&=&\frac{\kappa}{2} +D+\sqrt{\frac{\kappa^2}{4} +\kappa D}
\nonumber\\
 &=&\frac{\kappa}{2} +D+\frac{\kappa}{2} \left( 1+ \frac{2D}{\kappa}
-2\frac{D^2}{\kappa^2} +{\cal O}(3) \right) , \nonumber
\end{eqnarray}
so that to first order in $D$: $f(D) = $ $\kappa+2D$. From here we
obtain for (\ref{47})

\begin{eqnarray}
g(D)&=&\frac{\kappa}{2} , \label{48}\\
\text{meaning } g(D)&\sim &{\cal O}(D^0) .\nonumber
\end{eqnarray}
Using the value (\ref{48}) in eq. (\ref{46}) we arrive at

\begin{eqnarray}
0=&&\left[ (\partial_x \Phi)^2+ \left( 1+\frac{\kappa}{2D}
\right)(\partial_y \Phi)^2 \right] \partial_x^2 \Phi \nonumber\\
&+&\left[ \left( 1+\frac{\kappa}{2D} \right) (\partial_x \Phi)^2+
(\partial_y \Phi)^2 \right] \partial_y^2 \Phi \nonumber\\ 
&-&\frac{\kappa}{D} \partial_x \Phi\, \partial_y \Phi\, \partial_{xy}
\Phi . \label{49}
\end{eqnarray}
Following the strategies in QCD$_4$ it is convenient to rescale $x$
and $y$ in terms of dimensionless parameters:

\begin{equation}
x=R\,\bar{x}\quad ,\quad y=R^\alpha\,\bar{y} . \label{50}
\end{equation}
The authors of ref. \cite{7} supply arguments as to why the
transverse coordinate scales with $\alpha=\case{2}{3}$. Now, we try
the following ansatz: 

\begin{equation}
\Phi=\Phi^{(0)} +\frac{1}{R}\Phi^{(1)}+ \frac{1}{R^2} \Phi^{(2)}
+\dots\, .\label{51}
\end{equation}
Earlier we found $(2D)=((\partial_x \Phi)^2 +(\partial_y
\Phi)^2)^{1/2}$, so that

\begin{eqnarray}
\frac{1}{2D}&=& \left[ \frac{1}{R^2}(\partial_{\bar{x}}(\Phi^{(0)}
+\case{1}{R} \Phi^{(1)} +\dots))^2 \right. \nonumber\\
&&\left. \mbox{} +\frac{1}{R^{4/3}}(\partial_{\bar{y}} 
(\Phi^{(0)} +\case{1}{R} \Phi^{(1)} +\dots))^2 \right]^{-\case{1}{2}} 
\nonumber\\
&=&-\frac{R^{2/3}}{\partial_{\bar{y}} \Phi^{(0)}} +
\frac{(\partial_{\bar{x}} \Phi^{(0)})^2}{2(\partial_{\bar{y}}
\Phi^{(0)})^3} +{\cal O}(R^{-1/3}) , \label{52}\\
\text{with}&&\partial_{\bar{y}} \Phi^{(0)}<0 . \label{53}
\end{eqnarray}
Again, we skip the details of the calculation for the various partial
derivatives in (\ref{49}). Then we obtain the following differential
equation for the flux function in zeroth order ($\bar{x},\bar{y}
\equiv x,y$):

\begin{eqnarray}
0=&&\kappa\, (\partial_x^2 \Phi^{(0)}) (\partial_y \Phi^{(0)})
+\kappa \frac{(\partial_y^2 \Phi^{(0)})}{(\partial_y \Phi^{(0)})}
(\partial_x \Phi^{(0)})^2 \nonumber\\
&-& (\partial_y^2 \Phi^{(0)})(\partial_y \Phi^{(0)})^2 -2\kappa
(\partial_x \Phi^{(0)}) (\partial_{xy} \Phi^{(0)}). \label{54}
\end{eqnarray}
Another useful form of this equation can be obtained by
multiplication with $(\partial_y \Phi^{(0)})^2$:

\begin{equation}
0= \partial_y \left[ \frac{\kappa}{2} \left( \frac{\partial_x
\Phi^{(0)}}{\partial_y \Phi^{(0)}} \right)^2 +\partial_y \Phi^{(0)} 
\right] - \kappa \partial_x \left[ \frac{\partial_x
\Phi^{(0)}}{\partial_y \Phi^{(0)}} \right] . \label{55}
\end{equation}
We do not expect eqs. (\ref{54}) or (\ref{55}) to be soluble for
$\Phi^{(0)}(x,y)$ by the separation of variables method since there
is still the boundary condition (\ref{35}) to be taken into account.
Yet, as in QCD$_4$, there is hope for the existence of a separable
solution $y(x,\Phi)$. Hence our next step is to rewrite (\ref{54})
into a differential equation for $y$. Here are the rules governing
how to achieve this $(\Phi \equiv \Phi^{(0)})$:

\begin{eqnarray}
\Phi_y&=&y_\Phi^{-1}\, , \quad \Phi_x=-\frac{y_x}{y_\Phi}\, ,\quad
\Phi_{yy}=-\frac{y_{\Phi\Phi}}{y_\Phi^3} \label{56}\\
\Phi_{xx}&=&-\frac{y_{xx}}{y_\Phi}
-\frac{y_{\Phi\Phi}y_x^2}{y_\Phi^3}
+2\frac{y_{x\Phi}y_x}{y_\Phi^2}\nonumber\\ 
\Phi_{xy}&=&-\frac{y_{x\Phi}}{y_\Phi^2}+
\frac{y_{\Phi\Phi}y_x}{y_\Phi^3} . \nonumber
\end{eqnarray}
This set of partial derivatives enables us to rewrite (\ref{54}) in
the form

\begin{equation}
\frac{y_{\Phi\Phi}}{y_{\Phi}^5}-\kappa \frac{y_{xx}}{y_\Phi^2}=0 .
\label{57} 
\end{equation}
Here we can try the ansatz

\begin{equation}
y(x,\Phi)=X(x)\,F(\Phi) \label{58}
\end{equation}
and so obtain for (\ref{57}) instead

\begin{eqnarray}
\frac{F''}{FF'^3}&=&c , \label{59}\\
\kappa X''X^2&=&c , \label{60}
\end{eqnarray}
where $c$ is the separation constant. Integrating (\ref{59}) yields

\begin{equation}
\Phi=-\case{1}{6} c\frac{y^3}{X^3(x)} +k_1\frac{y}{X(x)}+ k_2 .
\label{61}
\end{equation}
With the aid of (\ref{35}) we find for $k_2$:

\begin{displaymath}
\Phi(y=0)=Q \quad : \quad k_2=Q \, .
\end{displaymath}
Since it is easier to work with $X(x)$ in the numerator of (\ref{61}) 
, we redefine $X \rightarrow \frac{1}{X}$ and so obtain

\begin{equation}
\Phi=Q(1-a_1yX(x)+a_3y^3X^3(x)) \label{62}
\end{equation}
\begin{equation}
\text{with new constants } \quad a_1=-\frac{k_1}{Q}\, ,\quad
a_3=-\frac{1}{6}\frac{c}{Q} . \label{63}
\end{equation}
Now recall that (i) we have to satisfy $\partial_y \Phi \leq 0$ in
the confining domain and (ii) we also want $\Phi$ to approach the
boundary $\Phi=0$ continuously: $\partial_y \Phi |_{y_{\text{b}}}=0$.
The first condition tells us that $a_1>0$ while the second implies
$a_3>0$. These considerations lead to the following two equations:

\begin{eqnarray}
0&=&\Phi(x,y_{\text{b}}(x)) \nonumber\\
 &=&Q(1-a_1\,y_{\text{b}}(x)X(x)+
a_3\,y_{\text{b}}^3(x)X^3(x)), \label{64}\\
0&=&\partial_y \Phi
|_{y=y_{\text{b}}} \nonumber\\
 &=&Q(-a_1X(x)+3a_3\,y^2_{\text{b}}X^3(x)). \label{65}
\end{eqnarray}
The last equation yields the explicit expression

\begin{equation}
y_{\text{b}}(x)=\pm \sqrt{\frac{a_1}{3a_3}} \frac{1}{X(x)} .
\label{66} 
\end{equation}
The two signs reflect the symmetry with respect to the $x$--axis.
Substituting (\ref{66}) into (\ref{64}) we obtain

\begin{equation}
a_3=\frac{4}{27} a_1^3 \, . \label{67}
\end{equation}
Thus, there is only one free parameter left. So far we have

\begin{eqnarray}
\Phi (x,y) &=& Q(1-\case{3}{2} b\, y X(x)+\case{1}{2} b^3 y^3 X^3(x))
\label{68}\\
y_{\text{b}}(x)&=&\pm \frac{1}{b} \frac{1}{X(x)} \, , \label{69}\\
\text{where }&&b=\case{2}{3} a_1\, ,\quad \text{or }\quad
a_3=\case{1}{2} b^3\,. \label{70}
\end{eqnarray}
What remains is an explicit solution for $X(x)$. Because of our
redefinition $X\rightarrow \frac{1}{X}$ the equation following from
(\ref{60}) is

\begin{equation}
\kappa \left( \frac{1}{X} \right)'' X^{-2}=c\, .\label{71}
\end{equation}
Introducing $X'=\sqrt{p}$ we can cast (\ref{71}) into the form

\begin{equation}
\frac{\text{d} p}{\text{d} X}=\frac{4p}{X} +6\frac{Q}{\kappa}b^3X^4\,
, \label{72} 
\end{equation}
which can be solved by

\begin{equation}
p(x)=\frac{6Qb^3}{\kappa} X^4 \left[ X-X_0 +\frac{\kappa p_0}{6QX_0^4 
b^3} \right]\, , \label{73}
\end{equation}
where $X_0=X(x=0)$ and $p_0=p(X_0)$. Among the three free parameters
$b,X_0,p_0$ we find, via equation (\ref{69}),

\begin{equation}
y_{\text{b}}(x)|_{x=0}=\frac{1}{bX_0}\, ,\label{74}
\end{equation}
that $b$ and $X_0$ merely rescale the size of the confinement domain.
Hence solution (\ref{73}) is essentially determined by $p_0$. Because
of

\begin{displaymath}
p_0=p(X_0)=p(X(x=0))=\left. \left(\frac{\text{d} X}{\text{d} x}
\right)^2 \right|_{x=0} 
\end{displaymath}
we see that $p_0$ is (a) related to the tangent of $\Phi$ in
$x$--direction at $x=0$,

\begin{equation}
\partial_x \Phi |_{x=0} =Q(-\case{3}{2} b\,yX'+\case{3}{2}
b^3y^3X^2X' )|_{x=0} \label{75}
\end{equation}
and (b) related to the shape of the confinement boundary at $x=0$:

\begin{equation}
\partial_x y_{\text{b}}(x)|_{x=0} = \left. -\frac{1}{bX^2}
X'\right|_{x=0}\, . \label{76}
\end{equation}
Now there are three different cases to be distinguished:

\begin{eqnarray}
\text{(1)}\quad&&p_{01}= \frac{6QX_0^5b^3}{\kappa} \nonumber\\
\text{(2)}\quad&&p_{02}= 0 \label{77}\\
\text{(3)}\quad&&p_{03}\neq p_{01}, p_{02}\, ,\quad
\text{arbitrary}\, . \nonumber
\end{eqnarray}
Case (1) has been treated in the literature \cite{6} and has the
advantage of being analytically soluble. We regard this solution as 
unphysical, since, looking at (\ref{76}), $y_{\text{b}}(x)$ does not 
behave smoothly at $x=0$; this solution never yields an extremum with 
regard to, e.g., the energy density, assuming an underlying
variational principle.
On the other hand, case (2) contains a physical, smooth boundary,
$\partial_x y_{\text{b}}|_{x=0}=0$, and is our preferred solution.
Its disadvantage is that it cannot be solved analytically. Case (3)
is neither analytically soluble nor is it physical and so will not be 
considered any further.

Without going into further details we now present our solution of
eqs. (\ref{54},\ref{55}) for the case (1):

\begin{eqnarray}
\Phi^{(0)}(x,y)&=&Q\left[ 1-2^{-2/3} \left( \frac{\kappa}{Q}
\right)^{1/3} \frac{y}{(a-x)^{2/3}} \right. \nonumber\\
&&\left.\, \quad +\frac{\kappa}{27Q} \frac{y^3}{(a-x)^2} \right]
\label{78}\\
y_{\text{b}}(x)&=&\pm \frac{3}{2^{1/3}} \left( \frac{Q}{\kappa}
\right)^{1/3} (a-x)^{2/3} \label{79}
\end{eqnarray}
(where we have switched back to our original coordinates $x,y
\rightarrow $ $\case{1}{R} x,\case{1}{R^{2/3}} y$). We have assumed
$x,y>0$; otherwise we would have to use moduli. Note the scaling
behavior of $y$ with a $\case{2}{3}$--power.

Finally we come to the calculation of the static potential. This is
achieved with the aid of the formula

\begin{eqnarray}
V_{\text{static}}&=&\int \text{d}^2x\, \int\limits_0^D E(D')\,
\text{d}D' \nonumber\\
 &=&\int \text{d}^2x\, \int\limits_0^D \text{d}D'\, \left[
\frac{\kappa}{2} +\sqrt{\frac{\kappa^2}{4} +\kappa D'} +D' \right]
\nonumber\\ 
 &=&\int \text{d}^2x\, \left( \frac{\kappa}{2}D +\frac{2}{3\kappa}
\sqrt{\frac{\kappa^2}{4} +\kappa D}^{\,3}-\frac{\kappa^2}{12}
+\frac{1}{2}D^2 \right) \nonumber\\
&=:& \int \text{d}^2x\, [\%]\, . \label{80}
\end{eqnarray}
The surface integration has to be performed over the confinement
region defined by (\ref{79}):

\begin{eqnarray}
V_{\text{static}}&=&\int \text{d}^2x\,
[\%]=\int\limits_{-a+\epsilon}^{a-\epsilon} \text{d}x
\int\limits_{-y_{\text{b}}(x)}^{y_{\text{b}}(x)} \text{d}y\, [\%]
\nonumber\\ 
&=&4\int\limits_{0}^{a-\epsilon} \text{d}x
\int\limits^{y_{\text{b}}(x)}_{0} \text{d}y\, [\%]\, . \nonumber
\end{eqnarray}
Restricting ourselves to weak fields, i.e., to large distances
(confinement or infrared domain), we approximate the integrand in
(\ref{80}) by

\begin{displaymath}
[\%]=\kappa D+D^2+\dots\, ,
\end{displaymath}
where $D=\case{1}{2} [(\partial_x \Phi)^2 +(\partial_y
\Phi)^2]^{1/2}$ in which we substitute the perturbative ansatz
(\ref{51}) and rescale according to (\ref{50}). An intermediate step
on our way to $V_{\text{static}}$ is

\begin{equation}
V_{\text{static}}=I_1+I_2\, , \label{81}
\end{equation}
where

\begin{eqnarray}
I_1&=&-2\kappa \int\limits_{0}^{a-\epsilon} \text{d}x
\int\limits_{0}^{y_{\text{b}}(x)} \text{d}y\, \left[ \partial_y
\Phi^{(0)} +\case{1}{R} \partial_y \Phi^{(1)} \right] \nonumber\\ 
 &=&-2\kappa \int\limits_{0}^{a-\epsilon} \text{d}x
\int\limits_{0}^{y_{\text{b}}(x)} \text{d}y\, \partial_y \Phi
\nonumber\\ 
&=&-2\kappa \int\limits_{0}^{a-\epsilon} \text{d}x \left[
\underbrace{\Phi|_{y_{\text{b}}(x)}}_{=0}
-\underbrace{\Phi|_{y=0}}_{=Q} \right] =2\kappa
Q\int\limits_{0}^{a-\epsilon} \text{d}x \nonumber\\
&=&\lim_{\epsilon \to 0} 2\kappa Q(a-\epsilon) =\kappa QR\, .
\label{82} 
\end{eqnarray}
So indeed we have linear confinement!

\noindent
The correction term $I_2$ turns out to be

\begin{equation}
I_2=\frac{12}{5} \sqrt[3]{\kappa Q^5}\, R^{1/3} -\frac{6}{35}
\sqrt[3]{\frac{Q^7}{\kappa}}\, R^{-1/3}\, . \label{83}
\end{equation}
Here, then, is our final expression [case(1)] for the static
potential of two oppositely charged massive test sources at large
distances: 

\begin{equation}
V_{\text{static}}=\kappa QR+\frac{12}{5} \sqrt[3]{\kappa Q^5}\,
R^{1/3} -\frac{6}{35} \sqrt[3]{\frac{Q^7}{\kappa}}\, R^{-1/3} +\dots 
\, .\label{84} 
\end{equation}
Note the linear rising with distance as it is familiar from QCD$_4$.

Now we turn to case (2) with $p_0=p_{02}=0$, which guarantees a
smooth boundary, $X'(x)|_{x=0}=0$. The equation to be integrated
follows from (\ref{73}):

\begin{equation}
\frac{\text{d} X}{\text{d} x} =\pm \sqrt{\frac{6b^3Q}{\kappa}}\,
X^2\sqrt{X-X_0}\, .\label{85}
\end{equation}
The solution is easily found:

\begin{equation}
x=\pm \sqrt{\frac{\kappa}{6Q(bX_0)^3}} \left[
\frac{\sqrt{\frac{X(x)}{X_0}-1}}{\frac{X(x)}{X_0}} + \arctan
\sqrt{\frac{X(x)}{X_0} -1} \right]. \label{86}
\end{equation}
This equation is transcendental with respect to $X(x)$, i.e., neither 
$X(x)$ nor $\Phi(x,y)$ can be written in an explicit analytical form.
It is, however, possible to find an approximate solution which is
sufficiently close to an exact solution and which maintains the
smooth boundary $y_{\text{b}}(x)$. Defining $t=X(x)/X_0$ it turns out
that $\sqrt{t-1}/t + \arctan \sqrt{t-1}$ can be excellently
approximated by $\arctan \sqrt{t^q -1}$ with $q\simeq 3.301\dots$,
whereby $q=3$ already represents a fairly good approximation. So we
can write instead of (\ref{86}):

\begin{eqnarray}
x&=&\pm \sqrt{\frac{\kappa}{6Q(bX_0)^3}} \arctan \sqrt{\left(
\frac{X(x)}{X_0} \right)^q -1} \label{87}\\
\text{or } X(x)&=&X_0 \left( 1+\tan^2 \left[
\sqrt{\frac{6Q(bX_0)^3}{\kappa}}\, x \right] \right)^{1/q}\, .
\label{88}
\end{eqnarray}
With the aid of eq. (\ref{69}), i.e., employing the condition

\begin{eqnarray}
0&=&y_{\text{b}}(x=\case{1}{2}) \nonumber\\
 &=&\pm \frac{1}{bX_0} \left( 1+\tan^2 \left[
\sqrt{\frac{6Q(bX_0)^3}{\kappa}}\, \case{1}{2} \right]
\right)^{-1/q}\, ,\nonumber\\
&&\text{ we obtain } \quad bX_0=\left( \frac{\kappa \pi^2}{6Q}
\right)^{1/3}\, .\nonumber
\end{eqnarray}
Hence our approximate solution for the flux function is given by
($x,y \rightarrow $ $x/R,y/R^{2/3}$):

\begin{eqnarray}
\Phi^{(0)}(x,y)=&Q&\left[ 1-\left( \frac{9\kappa \pi^2}{16Q}
\right)^{1/3} \frac{y}{R^{2/3}} \left(1+\tan^2 \case{\pi}{2}
\case{x}{a} \right)^{1/q} \right. \nonumber\\
 &&\left. \quad + \frac{\kappa \pi^2}{12Q} \frac{y^3}{R^2}
\left(1+\tan^2 \case{\pi}{2} \case{x}{a} \right)^{3/q} \right]
\label{89}
\end{eqnarray}
with the confining boundary

\begin{equation}
y_{\text{b}}(x)=2\left( \frac{3Q}{\kappa \pi^2} \right)^{1/3}
\frac{a^{2/3}}{\left(1+\tan^2 \case{\pi}{2} \case{x}{a}
\right)^{1/q}}\, .\label{90}
\end{equation}
Again, the scaling behavior $x\propto a$ and $y\propto a^{2/3}$ is
visible. 

With the aid of the flux function we now turn to the calculation of
the static potential. According to eqs. (\ref{81},\ref{82}) there is
no change in first (linear) order $\propto \kappa QR$; for the
calculation of this term the explicit form of $\Phi$ is not needed.
The correction term yields 

\begin{eqnarray}
I_2=-\int\limits_{0}^{a-\epsilon} \text{d}x
\int\limits_{0}^{y_{\text{b}}(x)} \text{d}y\,&& \left[ \kappa
\frac{(\partial_x \Phi^{(0)})^2}{\partial_y \Phi^{(0)}}
-(\partial_y \Phi^{(0)})^2 \right.\nonumber\\ 
&&\quad \left.-\frac{\kappa}{4}\frac{(\partial_x \Phi^{(0)})^4}
{(\partial_y \Phi^{(0)})^3}-(\partial_x \Phi^{(0)})^2
\right] \label{91} 
\end{eqnarray}
and after performing the various integrals we end up with

\begin{equation}
V_{\text{static}}=\kappa QR +2.17.. \sqrt[3]{\kappa Q^5}\, R^{1/3}
-0.36.. \sqrt[3]{\frac{Q^7}{\kappa}}\, R^{-1/3}\, . \label{92}
\end{equation}

\section{Short distance behavior}
Let us recall that in four dimensional space the static potential can
be evaluated by \cite{1} 

\begin{equation}
V_{\text{static}}=- (\text{extr}_\varphi W-\triangle
V_{\text{Coulomb}})\, ,\label{93}
\end{equation}
\begin{equation}
\text{where } W=\int \text{d}^3x\left[ \case{1}{8} b_0 (\bbox{\nabla}
\varphi)^2 \ln \left( \frac{(\bbox{\nabla} \varphi)^2}{\text{e}
\kappa^2} \right) -\varphi J_0 \right]. \label{94}
\end{equation}
By means of a cleverly chosen rescaling of coordinates, fields and
charges, Adler succeeds in splitting up the effective Lagrangian into
a classical part and extra terms due to quantum corrections
containing a dimensionless {\em running coupling} $\zeta(R)$. This
function goes to zero as $R$ approaches zero. Hence limiting oneself
to short distance behavior, $\varphi$ and $V_{\text{static}}$ can be
expanded in a perturbation series around the classical Coulomb
solution with $\zeta(R)$ as the small parameter.

\noindent
In QCD$_3$ we found for the effective action

\begin{eqnarray}
W&=& \int \text{d}^2 \left[ {\cal L}_{\text{eff}}(E) -\varphi J_0
\right] \nonumber\\
 &=& \int \text{d}^2 \left[ \case{1}{2} (\bbox{\nabla} \varphi)^2
\left( 1- \case{4}{3} \sqrt{\frac{\kappa}{E}} \right) -\varphi J_0
\right]\, .\label{95} 
\end{eqnarray}
Here it is obvious that the classical part $\propto (\bbox{\nabla}
\varphi)^2$ is already separated from the part containing the QCD
corrections. Hence, rescaling the various parameters does in no way
improve the situation. In fact the static potential approaches its
classical limit in a natural manner, since close to the test charges
we have $E\gg \kappa$. Thus, for short distances, $R \rightarrow 0$,
our zeroth order approximation is sufficiently descibed by
electrostatics: 

\begin{equation}
V_{\text{static}}=-\text{extr}_\varphi \int \text{d}^2x \left[
\case{1}{2} (\bbox{\nabla} \varphi)^2 -\varphi J_0 \right] +\triangle
V_{\text{c}}. \label{96}
\end{equation}
From (\ref{96}) we obtain the Poisson equation,

\begin{equation}
\bbox{\nabla}^2
\varphi=-J_0=-Q[\delta(\bbox{r}-a\bbox{\hat{x}})
-\delta(\bbox{r}+a\bbox{\hat{x}})]\,  , \label{97}
\end{equation}
which has the well--known solution

\begin{equation}
\varphi=-\frac{Q}{2\pi} \ln
\frac{|\bbox{r}-a\bbox{\hat{x}}|}{|\bbox{r}+a\bbox{\hat{x}}|} =-
\frac{Q}{2\pi} \ln \frac{\sqrt{(x-a)^2+y^2}}{\sqrt{(x+a)^2+y^2}}.
\label{98}
\end{equation}
The Coulomb counter term becomes

\begin{eqnarray}
\triangle V_{\text{c}}=\text{extr}_\varphi&& \left\{ \int \text{d}^2x
\left[ \case{1}{2} (\bbox{\nabla} \varphi)^2 -\varphi J_1 \right]
\right. \nonumber\\
 &&\quad +\left. \int \text{d}^2x \left[ \case{1}{2} (\bbox{\nabla}
\varphi)^2 -\varphi J_2 \right] \right\} \nonumber\\
J_1=Q&&\delta (\bbox{r}-a\bbox{\hat{x}})\, , \quad J_2=-Q\delta
(\bbox{r}+a\bbox{\hat{x}}) \nonumber
\end{eqnarray}
and the solutions of the corresponding field equations are

\begin{equation}
\varphi_1=-\frac{Q}{2\pi} \ln \lambda |\bbox{r}-a\bbox{\hat{x}}|\, ,
\quad \varphi_2=\frac{Q}{2\pi} \ln \lambda
|\bbox{r}+a\bbox{\hat{x}}|\, , \label{99}
\end{equation}
where the arbitrary parameter $\lambda$ has dimension [L]$^{-1}$.
Using the Poisson equation (\ref{97}) in the action we find for $W$
and similarly for $\triangle V_{\text{c}}$

\begin{eqnarray}
W&=&\int \text{d}^2x \left[ \case{1}{2} (\bbox{\nabla} \varphi)^2
-\varphi J_0 \right]\nonumber\\
& =&\int \text{d}^2x \left[ -\case{1}{2} \varphi \bbox{\nabla}^2
\varphi -\varphi J_0 \right] =-\case{1}{2} \int \text{d}^2x\, \varphi 
J_0\, . \nonumber 
\end{eqnarray}
Working with this expression for the action we obtain for
$V_{\text{static}}$ 

\begin{eqnarray}
V_{\text{static}}&=&\case{1}{2} \int \text{d}^2x\, \varphi J_0
-\case{1}{2} \int \text{d}^2x\, \varphi_1 J_1 -\case{1}{2} \int
\text{d}^2x\, \varphi_2 J_2 \nonumber\\
 &=& \frac{Q^2}{2\pi} \ln \lambda R\, . \label{100}
\end{eqnarray}
This, then, is the leading short distance behavior of the static
potential in QCD$_3$. In the sequel we will demonstrate that
expression (\ref{100}) also shows up when we now compute
$V_{\text{static}}$ by means of the formalism we developed for large
distances. To do so we have to return to the exact quasilinear,
second order differential equation (\ref{46}). Expanding the
coefficient function $g(D)$ in terms of $D$ near the sources where
$D\gg \kappa$, we obtain

\begin{equation}
g(D)=\case{1}{2} \sqrt{\kappa D} +\frac{\kappa}{4}+{\cal O}(D^{-1/2})
\, . \label{101}
\end{equation}
The relevant ratio $g(D)/D$ is thus of order ${\cal O}(D^{-1/2})$;
hence we are permitted to omit it for short distances, obtaining, not
surprisingly, Laplace's equation as approximation of eq. (\ref{46}):

\begin{equation}
(\partial_x^2+\partial_y^2) \Phi =0\, .\label{102}
\end{equation}
The well--known solutions satisfying the boundary conditions
(\ref{35}) are

\begin{equation}
\Phi (x,y)=\Phi_{\text{cl}}=\frac{Q}{\pi} \left[ \arctan
\frac{y}{x-a} -\arctan \frac{y}{x+a} \right]. \label{103}
\end{equation}
This expression for our flux function can be used to find

\begin{equation}
\bbox{D}=\frac{1}{2}\left( \begin{array}{c} \partial_y \Phi\\
-\partial_x \Phi \end{array} \right) =\frac{Q}{2\pi}  \left(
\frac{\bbox{\hat{r}}_1}{r_1}-\frac{\bbox{\hat{r}}_2}{r_2} \right).
\label{104} 
\end{equation}
Again, we come to the hardly surprising conclusion that the classical
linearization is sufficient for treating the short distance behavior
of two static color test sources.

At last we turn to the calculation of the static potential, which can
be achieved with the aid of formula (\ref{80}):

\begin{eqnarray}
V_{\text{static}}&=&\int \text{d}^2x [\case{1}{2}D^2+
\case{2}{3}\sqrt{\kappa} D^{3/2}+\case{\kappa}{2}D +{\cal
O}(D^{1/2})] \nonumber\\
 &=:& V_{\text{static}}^{D^2}+V_{\text{static}}^{V^{3/2}}+
V_{\text{static}}^{D} +\dots\, .\label{105}
\end{eqnarray}
Of course, $V_{\text{static}}^{D^2}$ yields the classical potential
and, with due consideration of the Coulomb counter terms, reads, as
before, 

\begin{equation}
V_{\text{static}}^{D^2} =\frac{Q^2}{2\pi} \ln \lambda R\, .\label{106}
\end{equation}
The other two remaining terms in (\ref{105}) allow us to augment the
classical potential by correction terms. Needless to say, the
respective integrations have to be performed with great care
\cite{7a}. The leading order contribution to the classical potential
comes from $V_{\text{static}}^{D^{3/2}}$, while $V_{\text{static}}^D$
provides us with a correction term $\propto R$. Our findings for the
potential of two static color charges at short distances can be
summarized in

\begin{eqnarray}
V_{\text{static}}=&&\frac{Q^2}{2\pi} \ln \lambda R \nonumber\\
&&+\frac{\sqrt{2}}{3} \left( \frac{\pi^2}{2} -\frac{\Psi'(1/4)}{4}
\right) \sqrt{\frac{\kappa Q^3}{\pi^3}}\, R^{1/2} \nonumber\\
&&+\frac{2G}{\pi} \kappa Q\, R\, , \label{107}
\end{eqnarray}
where $\Psi' (x)$ denotes the derivative of the psi function with
$\Psi'(1/4) \simeq$ 17.1973 and $G$ is Catalan's constant, $G=$
0.915965\dots . With these numbers we can rewrite (\ref{107}) in the
final form

\begin{equation}
V_{\text{static}}=\frac{Q^2}{2\pi} \ln \lambda R + 0.054..
\sqrt{\kappa Q^3}\, R^{1/2} +0.583.. \kappa Q\,R\, .\label{108}
\end{equation}
Hence in addition to the dominant classical potential, there exist
subdominant contributions behaving like $R^{1/2},R,\dots$ vanishing
as $R$ approaches zero. Equation (\ref{108}) should be read side by
side with Adler's formula (40) of ref. \cite{8}.

\section{Conclusion}
The results in this paper point to great similarities between QCD$_3$
and QCD$_4$. Despite major differences in physical and analytical
details we find that in both theories radiative corrections to
1--loop order spontaneously generate a gauge field vacuum condensate
leading directly to a confining theory. We find it interesting that
quantum contributions arise from the employment of purely classical
differential equations. In this way large distance as well as short
distance correction terms to the classical potentials were found.
Without overestimating the importance of low--dimensional field
theories, our calculation may lend some further insight into the
mechanisms of classical approximations of QCD$_4$.

\acknowledgements
One of us (W.D.) acknowledges various informative discussions with M.
Reuter. He also wants to thank the VW--Stiftung for the financial 
support during his Sabbatical in the U.S. where the present paper
took its final shape.


\begin{references}
\bibitem{1} S.L. Adler and T. Piran, Phys. Lett. {\bf 113 B}, (1982)
405; Phys. Lett. {\bf 117 B} (1982) 91

\bibitem{2} H. Pagels and E. Tomboulis, Nucl. Phys. {\bf B 143},
(1978) 485

\bibitem{3} H.D. Trottier, Phys. Rev. {\bf D 44}, (1991) 464

\bibitem{3a} M. Reuter and C. Wetterich, hep-th/9411227 v2

\bibitem{4} W. Dittrich and M. Reuter, {\em Effective Lagrangians in
QED}, in: Lecture Notes in Physics {\bf 220}, Springer--Verlag (1985)

\bibitem{5} J.M. Cornwall, Physica {\bf A 158}, (1989) 97

\bibitem{6} J. Frenkel and A.C. Silva Fo., Phys. Rev. {\bf D 33},
(1986) 2455

\bibitem{7} H. Lehmann and T.T. Wu, Nucl. Phys. {\bf B 237}, (1984)
205

\bibitem{7a} H. Gies, Phys. Lett. {\bf B 382}, (1996) 257

\bibitem{8} S.L. Adler, Nucl. Phys. {\bf B 217}, (1983) 381

\end{references}
\end{document}